\documentclass[preprint2]{aastex}

\shorttitle{GALAXY LIGHT CONCENTRATION}
\shortauthors{GRAHAM, TRUJILLO, CAON}
\slugcomment{To appear in AJ.}

\begin{document}

\input{psfig}

\title{GALAXY LIGHT CONCENTRATION. I. INDEX STABILITY AND THE CONNECTION WITH GALAXY STRUCTURE, DYNAMICS, AND SUPERMASSIVE BLACKHOLES} 
\author{Alister W. Graham, I. Trujillo, N. Caon}
\affil{Instituto de Astrof\'{\i}sica de Canarias, La Laguna,
E-38200, Tenerife, Spain}
\email{agraham@ll.iac.es, itc@ll.iac.es, ncaon@ll.iac.es}

\begin{abstract}

We explore the stability of different galaxy light concentration indices as a
function of the outermost observed galaxy radius.  With a series of analytical 
light-profile models,
we show mathematically how varying the radial extent to which one measures a
galaxy's light can strongly affect the derived galaxy concentration. 
The ``mean concentration index'', often used for parametrizing high-redshift 
galaxies, is shown to be horribly unstable, even when modelling one-component
systems such as Elliptical, dwarf Elliptical, and pure 
exponential disk galaxies.  The $C_{31}$ concentration index
performs considerably better but is also heavily dependent on the radial
extent, and hence exposure depth, of any given galaxy.  We show that the 
recently defined central concentration index is remarkably stable 
against changes to the outer radius and observational errors, 
and provides both a meaningful and reliable estimate of galaxy 
concentration.  The S\'ersic index 
$n$ from the $r^{1/n}$ models is shown to be monotonically related with the
central concentration of light, giving the index $n$ a second and perhaps more 
tangible meaning. 
With a sample of Elliptical and dwarf Elliptical galaxies we present 
correlations between the central light concentration and the 
global parameters: luminosity (Pearson's $r=-0.82$), 
effective radius ($r=0.67$), central surface brightness ($r=-0.88$), and velocity
dispersion ($r=0.80$).  The more massive Elliptical galaxies are shown to be
more centrally concentrated.   
We speculate that the physical mechanism behind the recently observed 
correlation between the central velocity dispersion (mass) of a 
galaxy and the mass of it's central supermassive blackhole may be connected 
with the central galaxy concentration.  That is, we hypothesize that 
it may not simply be the amount of mass in a galaxy, but rather how that 
mass is distributed, which controls the mass of the central blackhole. 

\end{abstract}


\section{Introduction}

Estimates to the central concentration of galaxy light for the 
parametrization of galaxies go back many decades (Morgan 1958, 
1959, 1962; subsequently known as the Yerkes system). 
Fraser (1972) made this method of classification 
quantitative with the introduction of the concentration 
indices $C_{21}$ and $C_{32}$, defined as the ratio of radii
that contain 50\% and 25\%, and 75\% and 50\%, of the total 
(asymptotic) galaxy luminosity.  de Vaucouleurs (1977) expanded 
on this with the introduction of the more commonly used $C_{31}$ index 
(see, for e.g., Kent 1985; Gavazzi, Garilli, \& Boselli 1990; Moriondo 
et al.\ 1999). 

Okamura, Kodaira, \& Watanabe (1984) explored other fractional ratios 
and also introduced what they called the ``mean concentration index''. 
This was a ratio of the luminous flux enclosed by two different radii
rather than a ratio of radii (see section~\ref{oka}).  
Using an outer isophotal level of 26 $V$-mag arcsec$^{-2}$, 
Okamura et al.\ described how these indices are dependent on the 
intrinsic (or mean) surface brightness of the individual galaxies.  
At least for Elliptical galaxies, or rather, one-component galaxies 
or bulges -- which is what we wish to explore here -- this can be 
directly translated into a dependency on the number of effective 
radii one samples.  It is desirable to perform this translation 
because one replaces a dependency on two quantities, namely, 
the intrinsic surface brightness of each galaxy and the faintest surface 
brightness level observed, with one quantity (i.e.\ the number
of scale-lengths or effective radii observed).  It is this dependency 
which we wish to explore. 
Ideally, one would like to have a concentration index which is as 
independent as possible on this quantity (which is dependent on, 
amongst other things, the  observational exposure details).  

We use the S\'ersic (1968) $r^{1/n}$ law as a model-dependent way 
to illustrate the various concentration indices presented in Section 2.  
(The concentration indices, however, can be measured independently 
of this, and any, model.)  
All the concentration indices are dependent in some way on the 
extent to which the galaxy radial profile is sampled, and this 
dependency is revealed in Section 3.  
Truncating analytical light-profile models, we perform a comparative 
study of three different concentration indices.  Additionally, 
using a range of observed Elliptical galaxy profiles obtained 
from deep exposures, we again explore the stability of the concentration 
indices. 

In Section 4 we investigate the relationship between the galaxy light 
concentration and the other physical galaxy properties such as: 
luminosity, effective radius, central surface brightness, and velocity 
dispersion. 
Under the assumption that Elliptical galaxies are homologous systems, 
that is, assuming they all obey the $r^{1/4}$ law, the concentration 
index of every Elliptical galaxy should be the same.  However, research 
over the last decade has shown that Elliptical galaxies are not 
homologous and we present, in some cases, the first ever correlations 
between galaxy concentration and the global properties of Elliptical and 
dwarf Elliptical galaxies.

\section{Concentration indices}

We use the S\'ersic $r^{1/n}$ law as a mathematical means to illustrate
the various concentration indices.
The $r^{1/n}$
luminosity-profile model has been shown to provide a good representation to
the distribution of light in both Elliptical galaxies (including the dwarf
Ellipticals) and the bulges of Spiral galaxies (Caon, Capaccioli, \& D'Onofrio
1993;  D'Onofrio, Capaccioli, \& Caon 1994; Young \& Currie 1994;  Andredakis,
Peletier, \& Balcells 1995), and is also appropriate for describing a Spiral 
galaxy's disk.  
Although it's appearance in the literature is becoming more frequent, 
we summarize it's form below and introduce the parameters we will 
subsequently use throughout the paper. 

The de Vaucouleurs (1948, 1959) $r^{1/4}$ radial intensity profile  $I(r)$ was
generalized by S\'ersic (1968) to give the $r^{1/n}$ law where 
\begin{equation} 
I(r)=I(0)\exp^{-b_n(r/r_e)^{1/n}}.
\end{equation}
$I(0)$ is the central intensity and $r_e$ the effective radius enclosing half
of the flux from the model light profile.   The quantity $b_n$ is a function of
the shape parameter $n$ -- which defines the global curvature in the luminosity
profile -- and is obtained from solving the expression $\Gamma (2n)$$=$$2\gamma
(2n,b_n)$,  where $\Gamma(a)$ and $\gamma(a,x)$ are respectively the gamma
function and the  incomplete gamma function.   
The total luminosity, $L_T$, associated with an $r^{1/n}$ profile 
that extends to infinity can be written as
\begin{equation}
L_T=I(0)r_{e,{\rm mod}}^2\frac{2\pi n}{b_n^{2n}}\Gamma(2n), 
\label{eqlum1}
\end{equation}
where $r_{e,{\rm mod}}$ is the effective half-light radius 
of the model.  For elliptical (that is, non-spherical) galaxies, 
$r_{e,{\rm mod}}^2 = r_{e,{\rm maj}}^2(1-\epsilon)$, where 
$\epsilon$ is the ellipticity of the galaxy and $r_{e,{\rm maj}}$ 
is the major-axis half-light radius. 
The outer radius to which one (reliably) measures a galaxy is of course 
a function of exposure time, telescope aperture, etc.  
Denoting this outer finite (or final) radius as $r_{\rm fin}$, the 
luminosity enclosed by this radius is 
\begin{equation}
L(r_{\rm fin})=I(0)r_{e,{\rm mod}}^2\frac{2\pi n}{b_n^{2n}}
\gamma\left(2n,b_n\left(\frac{r_{\rm fin}}{r_{e,{\rm mod}}}\right)^{1/n}\right).
\label{eqlum2}
\end{equation}

Having introduced a model which can be used to represent the observed 
range of structural profile shapes in bulges, we can now proceed to 
describe the degree of 
concentration of light in these systems.  The following definitions 
for the various concentration indices can all be applied without reference
to the above (or any) light-profile model; but it is of course 
insightful to use a parametrized model. 
We give below definitions or procedures to obtain the various concentration
indices as a function of the outermost observed radius.

\subsection{The concentration index $C_{31}$} 

Due to the popular use of $C_{31}$ over $C_{32}$ and $C_{21}$, we will focus on
this particular concentration index related to the ratio of radii.  The
$C_{31}$ index can either be obtained directly from the image,  or using
equation~\ref{eqlum1} and \ref{eqlum2} when the observed  luminosity profile is
well fitted with an $r^{1/n}$ model.   Because light-profile models can
first be convolved with the relevant point spread function before fitting 
to an observed profile, the latter approach has the strong  advantage that
corrections for seeing are already taken into account.  Dividing
equation~\ref{eqlum2} by equation~\ref{eqlum1} gives the  fractional
luminosity, denoted here by $x$, contained within the outermost observed 
radius $r_{\rm fin}$.  One then solves the ratio of  equation~\ref{eqlum2} 
and equation~\ref{eqlum1} for the (new) values of $r_{\rm fin}/r_{e,{\rm mod}}$ 
that give a fractional luminosity of $0.75x$ and $0.25x$.   The results of 
doing this are shown in Section 3.

\subsection{The mean concentration index \label{oka}}

The fundamental parameter in the classification system of 
Abraham et al.\ (1994) is basically the concentration index 
defined by Okamura et al.\ (1984) and Doi et al.\ (1993).  
It is a luminosity ratio between the flux enclosed by some inner radii 
and the outer-most radii, and has been parametrized in Abraham 
et al.\ (1994) such that 
\begin{equation}
C(\alpha)=\frac{\sum_{i,j\in E(\alpha)}I_{ij}}{\sum_{i,j\in E(1)}I_{ij}}.
\label{eqAbr1}
\end{equation}
Here, $I_{ij}$ represents the intensity in the pixel $(i,j)$, and
$E(\alpha)$ denotes some inner radius which is $\alpha$ ($0<\alpha<1$) 
times the outer radius which has been normalized to 1.  Following 
Doi et al.\ (1993) Abraham et al.\ (1994) used a value of 0.3 for 
$\alpha$.  In what follows in Section 3, we will use a value of 
$\alpha=1/3$.  

Trujillo et al.\ (2001b) 
noted that as the exposure depth increases and the outer radius 
therefore increases, this definition of light 
concentration loses its significance and tends to a value of 1 
for all galaxies. 
For the S\'ersic models, this concentration index can be written 
as a function of the outer radius $r_{\rm fin}$ such that 
\begin{equation}
C(\alpha)=\frac{\gamma\left[2n,b_n\left(\alpha \frac{r_{\rm fin}}
{r_{e,{\rm mod}}}\right)^{1/n}\right]}
{\gamma\left[2n,b_n\left(\frac{r_{\rm fin}}
{r_{e,{\rm mod}}}\right)^{1/n}\right]}.
\label{eqAbr2}
\end{equation}

\subsection{A new concentration index}

\begin{figure}
\plotone{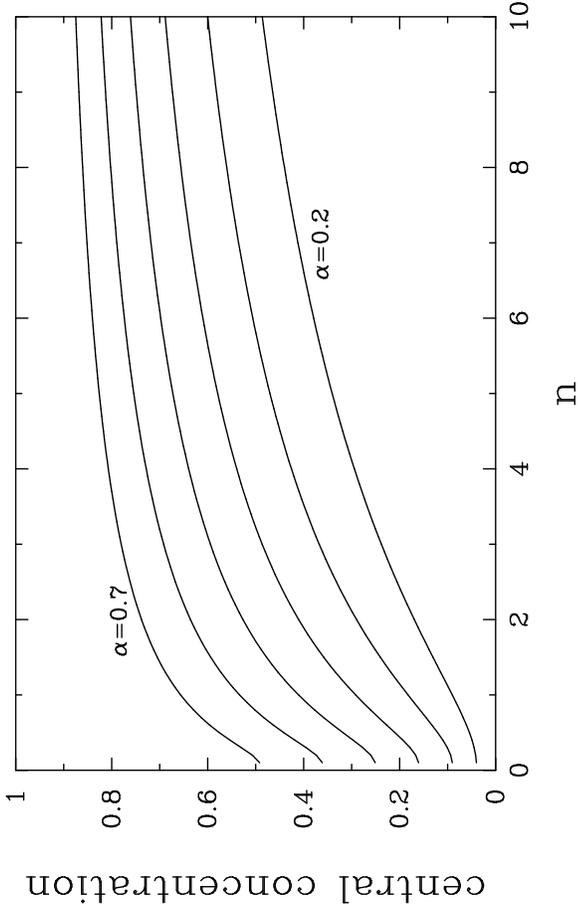}
\caption{
The central galaxy concentration index as defined in equation~\ref{eqTGC1}
and equation~\ref{eqTGC2} is shown as a function of the S\'ersic shape
parameter $n$, for different values of $\alpha$.
(Extension of Figure 3 from Trujillo et al.\ (2001b) who plotted values
of $\alpha$ equal to 0.3 and 0.5.)
}
\label{fig1}
\end{figure}

\begin{figure}
\plotone{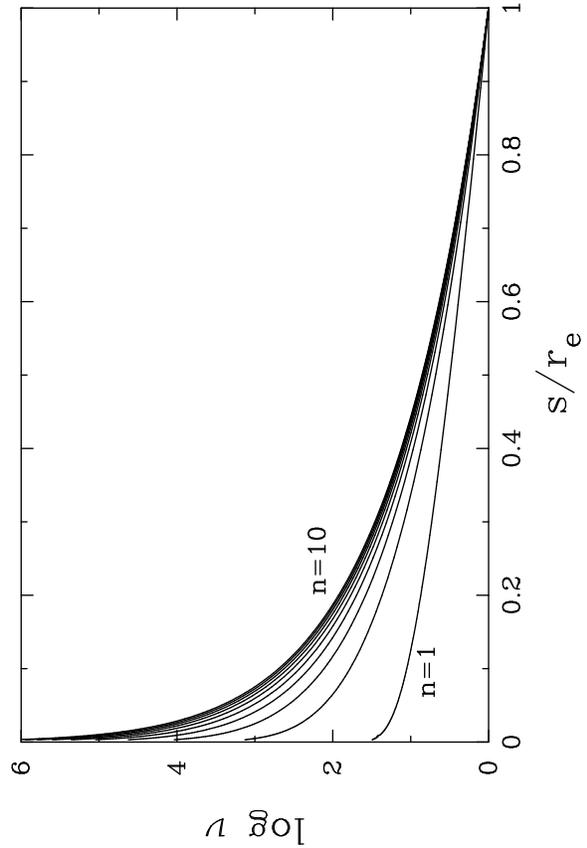}
\caption{
The deprojected (spatial) luminosity-density profiles for a range
of S\'ersic $r^{1/n}$ models are shown as a function of the deprojected
radius $s$, normalized at 1 $r_e$ from the projected $r^{1/n}$ model.  The
increased central concentration of these models with $n$ is strong and obvious.
}
\label{fig2}
\end{figure}

We describe here a `third galaxy concentration' (TGC) index defined in Trujillo
et al.\ (2001b) as 
\begin{equation} 
TGC(\alpha)=\frac{\sum_{i,j\in E(\alpha r_e)}I_{ij}}{\sum_{i,j\in E(r_e)}I_{ij}}. 
\label{eqTGC1} 
\end{equation}  
Here, $E(r_e)$ means the isophote which encloses half of the total light 
of the galaxy\footnote{In practice, $r_e$ is the observed half-light radius
$r_{e,{\rm obs}}$ and not $r_{e,{\rm mod}}$.}, and $E(\alpha r_e)$ is the
isophote at a radius $\alpha$ ($0<\alpha<1$) times $r_e$.  Again, this is a
flux ratio.  For a S\'ersic profile which extends to infinity, 
\begin{equation} 
TGC(\alpha)=\frac{\gamma(2n,b_n\alpha^{1/n})}{\gamma(2n,b_n)}.
\label{eqTGC2} \end{equation} 
For a range of different values of $\alpha$, this concentration index is 
shown in 
Figure~\ref{fig1} to increase  monotonically with the value of $n$, revealing a
relation between the central galaxy concentration of light and the global
galaxy structure as defined by the shape parameter $n$ (see Trujillo et al.\ 
2001b).  That is, the shape
parameter $n$ can be thought of as more than just a parameter that describes
the curvature of the light profile, but has the additional physical meaning
that it describes the degree of central concentration of at least luminous 
matter in Elliptical galaxies and bulges. 
This relation between the central luminous concentration and the global 
structure of the galaxy can be made even more succinctly by deprojecting the 
various $r^{1/n}$ profile models to obtain their spatial luminosity density 
profiles (Ciotti 1991) for different values of $n$ (Figure~\ref{fig2}).  
One can immediately see the increasingly dramatic rise in central 
density\footnote{For $n>1$ the central density is actually 
infinite, occurring at a singularity (Ciotti 1991).} with increasing $n$. 

For larger values of $\alpha$ (e.g.\ $>$0.5) the TGC index loses its 
ability to clearly distinguish galaxies with different structural  profile 
shapes (i.e.\ different $n$).  Low values for $\alpha$,  such as  0.2, provide
a good range to the TGC index for different $n$,  but in reality  are not so
practical (especially when dealing with  high-redshift galaxies).   We
therefore propose to use $\alpha =1/3$ (keeping the 3:1 ratio of the 
previous concentration indices). 

Equation~\ref{eqlum2} shows how reducing the outer observed radius  $r_{\rm
fin}$ of an $r^{1/n}$ profile reduces the observed galaxy  luminosity.  This in
turn reduces the observed effective half-light radius from $r_{e,{\rm mod}}$ to
$r_{e,{\rm obs}}$ (see Trujillo et al.\ 2001b, their Figure 9).   As a result,
the TGC index depends on the outer radius in the  following way
\begin{equation} TGC(\alpha)=\frac{\gamma\left[2n,b_n\left(\alpha
\frac{r_{e,{\rm obs}}} {r_{e,{\rm mod}}}\right)^{1/n}\right]}
{\gamma\left[2n,b_n\left(\frac{r_{e,{\rm obs}}} {r_{e,{\rm
mod}}}\right)^{1/n}\right]}.  \label{eqTGC3} \end{equation} where $r_{e,{\rm
obs}}/r_{e,{\rm mod}}$ must first be computed  given $r_{\rm fin}/r_{e,{\rm
mod}}$ (Trujillo et al.\ 2001b, their section 5.2).

\section{Stability Analysis of the various concentration indices}

\begin{figure}
\plotone{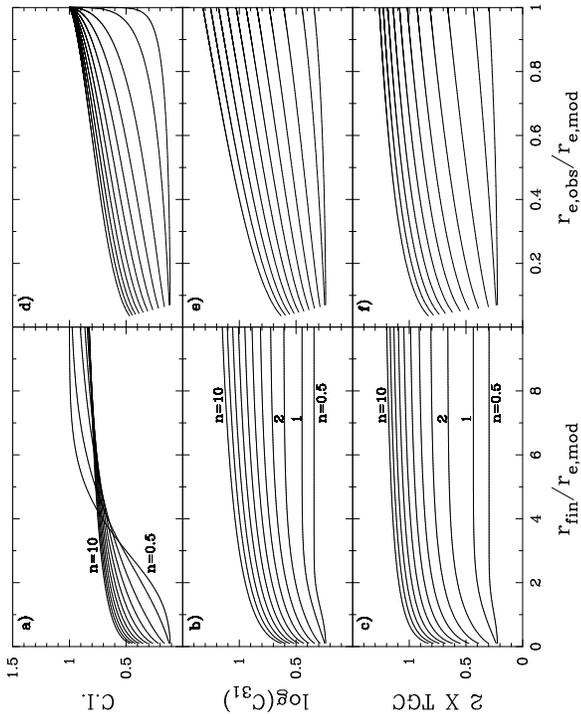}
\caption{Various concentration indices have been plotted, for different
galaxy profile shapes $n$, as a function of the outer radius
$r_{\rm fin}$ used to compute each index (see text for details).
Panel a) presents the concentration index as defined by Abraham et
al.\ (1994).  Panel b) presents the logarithm of the concentration index
defined be de Vaucouleurs (1977).  Panel c) presents the concentration
index defined by Trujillo et al.\ (2001b).
Panel d) to f) show the various indices as a function of the observed
half-light radius, which changes with $r_{\rm fin}$.
The value of $\alpha$ used to compute $C_{31}$ and $TGC$ has been
set to 1/3.
The TGC index is clearly the most stable of the three, while the index
in panel a) is horrendously unstable.
}
\label{fig3}
\end{figure}

In Figure~\ref{fig3} we have shown how the above three concentration  indices
vary as the outer final radius is varied.  One can see that  the commonly used
concentrations indices do not  simply reveal the luminous structural
concentration one hopes to measure but can be heavily biased by the radial
extent used to compute them.  Consequently, concentration indices derived for
exactly the same  galaxies observed with first a `shallow' and then a `deep'
exposure  will be different.   Clearly the ``mean concentration index''
(Section~\ref{oka}) is unreliable by itself to  provide any kind of 
meaningful galaxy 
classification.   Okamura et al.\ (1984) recognized this short-coming and,
despite  their hope to define a single fundamental structural quantity,  had to
resort to the introduction of an additional parameter, namely, the mean surface
brightness of each galaxy.  Even then they  still had to  acknowledge that the
range of different global profile shapes (due also to the different bulge/disk
combinations which galaxies  possess) introduced additional scatter which left
the previous two parameters still unable to provide an accurate classification
of galaxy types along the morphological sequence (their Figure 3).   They were
however able to accurately (85\%) classify galaxies as either early-type
(E-S0/a) or late-type (Sb-Im).   This broad categorical restriction was
confirmed by Doi et al.\ (1992), Doi, Fukugita, \& Okamura  (1993), and then
again by  Abraham et al.\ (1994) who explicitly dealt with ellipticity and 
extended the formalism to the study of small and faint images  where individual
pixel information is important.  Doi et al.\ (1993) also showed that the effects
of seeing can result  in one not even being able to distinguish between these
two broad categories when a small number of scale-lengths are sampled.  Most
recently, Bershady, Jangren, \& Conselice  (2000)\footnote{Following Kent
(1985), Bershady et al.\ (2000) used a ratio of radii for their concentration
index $C$, such that  $C=5\log [r(80\%)/r(20\%)]$.}  have introduced a third
parameter, namely color -- which Hubble  (1936) noted was correlated with
morphological type -- that  has enabled them to separate galaxies into three
classes:  early, intermediate, and late-type. 

While the ``mean concentration index'' in combination with other galaxy 
parameters may be able to broadly categorize galaxies, it should 
clearly not be used as a tracer of the concentration of the 
luminous matter in galaxies. 
Its value depends sensitively on the depth of the image, and is 
therefore not closely related to any underlying physical 
property of the galaxy.  Even for the same galaxy cluster, 
where the exposure details are the same, the situation is a mess.  
Figure~\ref{fig3} reveals that this index cannot even 
distinguish between a pure exponential disk (for e.g.\ an Sd galaxy) 
observed to three effective radii and a giant $r^{1/4}$ elliptical 
galaxy observed to one effective radii. 

The original concentration index defined as the ratio between 
radii enclosing different fractions (or quartiles) of the total 
galaxy luminosity (Fraser 1972; de Vaucouleurs \& Aguero 1973; 
Fraser 1977; de Vaucouleurs 1977) is significantly more stable 
than the ``mean concentration index''.  This is because this 
index is less affected by where the galaxy profiles are taken to 
terminate.  
Although, it should be noted that this index is 
noticeably less well behaved when one removes the logarithm used 
in Figure~\ref{fig3}.  On the other hand, the TGC index appears 
to be very stable. 

\begin{figure}
\plotone{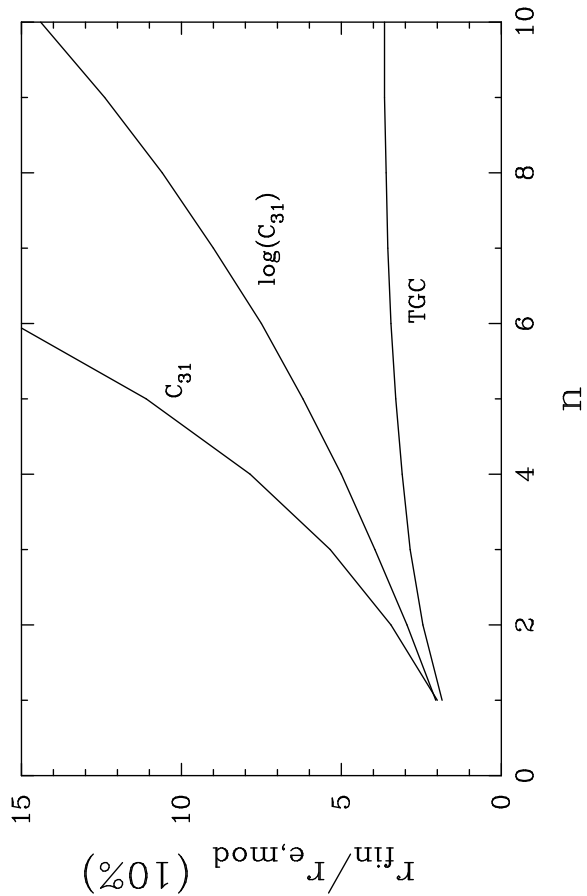}
\caption{Stability analysis of the the $C_{31}$ concentration index
and the TGC index.  The radius $r_{\rm fin}$ (normalized by the
effective radius $r_{\rm e,mod}$ of each $r^{1/n}$ model) where the
concentration indices get within 10\% of their maximum value (which occurs
at an infinite radial extent) is shown as a function of $n$. The TGC
index is clearly more stable with radius, and hence exposure depth,
than the $C_{31}$ index.
}
\label{fig4}
\end{figure}

In order to compare the $C_{31}$ index with the TGC index, we  have plotted in
Figure~\ref{fig4} the radii at which the two  indices get within 10\% of their
asymptotic value, which occurs at an infinite radial extent. One can
see that the TGC index acquires this level at a reasonable number of effective
radii. The $C_{31}$ index performs significantly worse.  
We therefore conclude that of the various concentration 
indices, the ``mean concentration index'' should not be used for the  study of
one-component systems (i.e.\ Elliptical galaxies, dwarf Elliptical galaxies,
Spiral galaxy bulges, pure  exponential disks).  It's application to
two-component systems  may be even less appropriate and will be studied in a
forth-coming  paper.  The TGC index should be used in preference to the
$C_{31}$ index.  The relative independence of the TGC index on the observed
radial  extent, and therefore on the individual intrinsic galaxy surface 
brightness and exposure details, may mean that this single parameter can be
used on its own to quantify, at least one-component systems, without the need 
to obtain and calibrate surface brightness levels and colors; 
that is, even non-photometric data could be analyzed.

\subsection{Tests with observed galaxy profiles}

In the above section we have analyzed the stability of the various 
concentration indices against truncations in the radial extent of 
model S\'ersic light-profiles.
We now perform one of what is no doubt many possible tests which has no 
dependency on the S\'ersic model and uses real galaxy profile data 
containing noise, sky-subtraction errors, and possible deviations from 
perfect $r^{1/n}$ models. 

In practice, to determine the total galaxy light attempts are made 
to account for the fraction of 
light which may reside outside the last measured isophote.
In order to do this, the growth curve is normally extrapolated by one of 
a variety of techniques, such as a convenient mathematical function, or 
fitting some ad-hoc model, or even sometimes by eye.  
The extrapolation term, and the corresponding uncertainty, depend on a 
variety of factors, such as the shape of the light profile, the depth of
the galaxian image, the accuracy of the sky-background subtraction, the
particular way the extrapolation is computed, and so on. 
For this reason, it is of interest to compare (in a model-independent way) 
the measured concentration indices using observed profiles, and to check which 
index is least sensitive (i.e.\ more robust) to the above potential sources 
of error in estimating the total galaxy light.  
The ``mean concentration index'' will not be computed here as 
it has already been shown to be a poor estimator of concentration, 
equal to 1 for any profile measured to a large radial extent. 

We have analyzed the Virgo galaxy profiles presented in 
Caon et al.\ (1993, 1994) through the following experiment:
\begin{enumerate}
\item The galaxy growth curve was computed, out to the outermost measured
radius, using the observed major-axis light profile and the ellipticity
and position angle profiles.
The total $B$-band magnitude was derived by extrapolating the growth curve
to infinity (see the above two references) to give $m_{\rm B}$.  
\item Next, a variable magnitude $\Delta m$ was added to $m_{\rm B}$ to
give $m_{\rm tot}$.
%
%
Negative values of $\Delta m$ correspond to changes in the magnitude due to
an overestimate of the total galaxy luminosity; positive values may represent 
truncations in the light profile, or represent magnitudes within some
isophotal threshold (such as $V_{26}$), or simply account for errors from the
true total galaxy magnitude.
\item From the value of $m_{\rm tot}$ we then derived the effective half-light
radius $r_{\rm eff}$ and computed the TGC index.  We also derived $r_{25}$
and $r_{75}$, the radii at which the enclosed luminosity is 25\% and 75\% of
the total luminosity, and determined the value of $C_{31}$.
Both concentration indices were derived independently of any profile model.
\item Steps 2 and 3 were repeated for $\Delta m$ ranging from $-0.2$ to 0.4.
\end{enumerate}
By plotting these model-independently derived TGC and $C_{31}$ indices 
as a function of $\Delta m$, we were able to see how they reacted to 
uncertainties/changes/errors in the total galaxy magnitude. 

\begin{figure}
\plotone{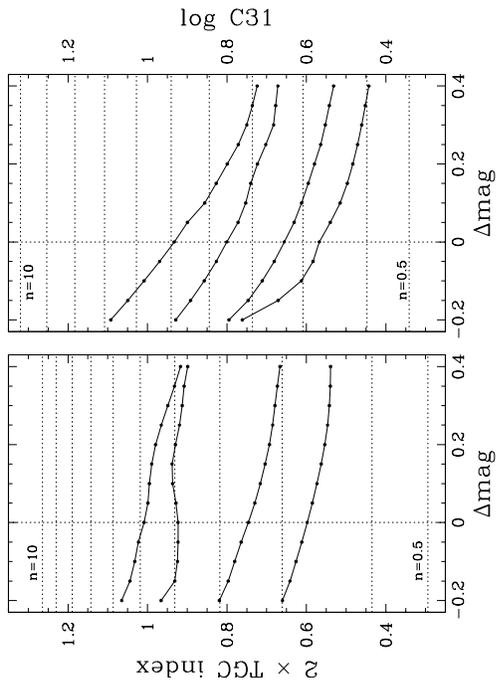}
\caption{TGC and $C_{31}$ indices (computed independently
of any S\'ersic $r^{1/n}$ model) for a sample of real galaxy 
profiles spanning a range of structural shapes (i.e.\ $n$).  
The light-profile data was taken from Caon et al.\ (1990) and (1994)
for NGC~4623, NGC~1379, NGC~4636, and NGC~4365, in order of 
increasing galaxy light concentration. 
The influence on the concentration indices by supplementing the 
observed galaxy magnitude within some outer radius with $\Delta m$ is 
revealed (see text for details).  The term $\Delta m$ can represent 
the magnitude beyond 
the outer radius which was missed, or represent truncations to an inner
isophotal level, or one of a number of sources of observational error.
Dotted horizontal lines mark the values of the TGC and $C_{31}$ 
indices corresponding to infinite S\'ersic models with $n=0.5,1,2 {\ldots} 10$.
}
\label{fig5}
\end{figure}

For a given variation in galaxy profile shape, and therefore galaxy 
concentration, $C_{31}$ may change its value by $x$\% while the TGC 
index changes its value by $y$\%, or vice-versa.  
Therefore the stability of different indices can not be compared in 
terms of their percentage changes, as different percentage changes
may accurately reflect identical changes in galaxy structure.  
The sensitivity of each index to measurement errors should therefore 
perhaps be viewed in the light of the implied changes in galaxy structure.  
Figure~\ref{fig5} shows the derived TGC 
and $C_{31}$ indices plotted against $\Delta m$ for a random sub-sample 
of galaxies possessing a range of S\'ersic indices $n$.  
It clearly shows that the TGC index is more stable than $C_{31}$ 
because, for the same change in $\Delta m$, the TGC index spans a smaller 
interval in $\Delta n$, that is, a smaller change in galaxy structure.  
In other words, if one used the model-independent derivations for the TGC 
and $C_{31}$ indices as a way to estimate the true galaxy structure as 
represented by $n$, the TGC index would give a more stable estimate which is 
less prone to uncertainties in $m_{tot}$ than the index $C_{31}$ is. 

\begin{figure}
\plotone{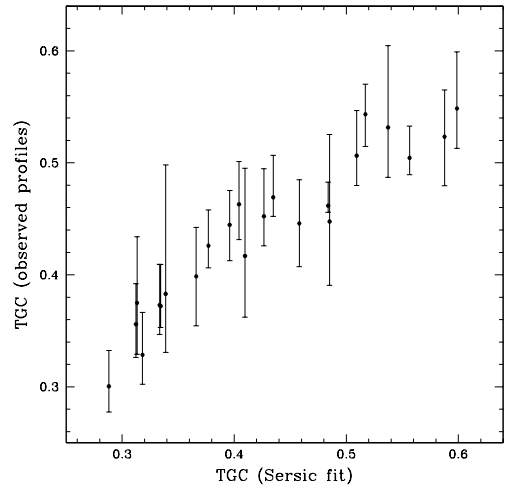}
\caption{The stability of the TGC index has been explored with real galaxy 
data by deriving 
its (model-independent) value when an error of $\pm$0.2 mag is added to the 
(model-independent) estimated total galaxy magnitude.  The $x$-axis is the 
TGC index derived 
from the best-fitting S\'ersic model.  Only galaxies with major-axis 
fits which were marked as `Good' (that is, not those whose quality was 
marked as `Fair' or `Poor') from Table 2 of Caon et al.\ (1993) have been used. 
}
\label{fig6}
\end{figure}

\begin{figure}
\plotone{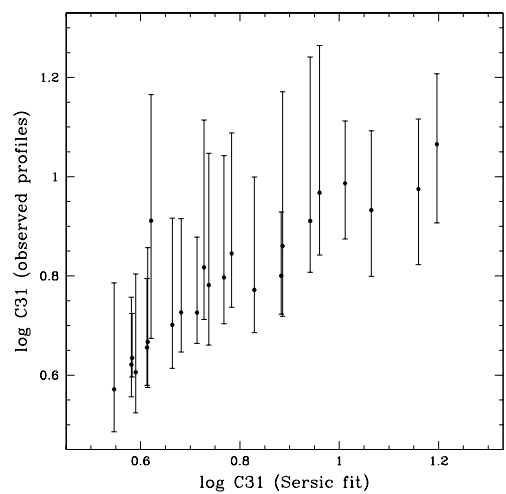}
\caption{Same as Figure~\ref{fig6} but for the $C_{31}$ index.}
\label{fig7}
\end{figure}

The above point is dramatically illustrated in Figure~\ref{fig6} and 
Figure~\ref{fig7}.  Figure~\ref{fig6} shows the TGC index derived from the 
best fitting S\`ersic model to the observed galaxy light profile and the 
TGC index derived directly from the light profile data itself, with no 
dependence on a $r^{1/n}$ model.  In the model-independent case, 
the estimated total galaxy magnitude was
then increased and decreased by $\Delta m$=0.2 mag (i.e.\ spanning a 
range of 0.4 mag) and the TGC index re-computed.  
The first thing one notes from Figure~\ref{fig6} is that the model-dependent
and model-independent values agree reasonably well with each other.
The second point, which receives emphasis when one simultaneously 
considers Figure~\ref{fig7}, is that the range in values for the TGC
index is quite well constrained when one varies the total galaxy flux
by nearly some 50\%.  The $C_{31}$ index does not 
behave anywhere near as well (Figure~\ref{fig7}).  
It can be clearly seen to be far less stable to errors in the total 
galaxy flux than what the TGC index is.

\section{Correlations between galaxy concentration, structure,
and dynamics}

The concentration index, in its various guises, has been shown to correlate,
albeit sometimes poorly, with galaxy morphological type; indeed, Doi et al.\
(1993) suggested it be used, together with the observed mean surface
brightness, as a means to morphologically classify different galaxy types 
(see also Bershady et al.\ 2000 and references within).  We show for the 
first time in the following 
subsections that the TGC index, at least for the family of Elliptical 
galaxies, is strongly correlated with all the fundamental galaxy
parameters.  The concentration appears to not only reflect the general
morphological structure, but is intimately related with the total luminosity,
size, brightness, and central velocity dispersion of a galaxy. 

One may ask, ``But isn't the TGC index merely another way of expressing 
the exponent $n$?'' Or, is this central concentration index (a quantity 
which can be measured independently of any model or value of n) a 
fundamental quantity which is intimately linked with the nature and 
evolution of Elliptical galaxies. 
What are the fundamental quantities which should be plotted against
each other to gain insight into the nature of Elliptical galaxies?  
At this point we don't know.  We therefore show, for 
a sample of Elliptical galaxies, correlations between
concentration and: luminosity, surface-brightness, scale-size. and 
velocity dispersion. 
What past correlations with n actually mean are somewhat vague.
Only when you understand what the correlations mean are you
able to say something about the physical behavior of galaxies. 
What we shall see below is that the larger, more luminous and 
massive galaxies are more centrally concentrated.

\subsection{Luminosity \label{ozzy}}

The previous analysis, in particular Figure~\ref{fig1}, reveals that the
central luminous concentration in Elliptical galaxies, and also Spiral 
galaxy bulges, must be related to their global luminous structure. 
This is because the more luminous galaxies and 
bulges are known to possess larger values of $n$ (Caon et al.\ 
1993; Young \& Currie 1994, 1995; Andredakis et al.\ 1995; Jerjen 
\& Binggeli 1997; Graham 2001) and consequently they must also have 
higher central concentrations of 
light than the less luminous bulges. Figure~\ref{fig8} shows that this is
indeed the case.  Here, the TGC index has been plotted against the bulge
luminosity from a sample of Virgo dwarf Elliptical galaxies (Jerjen, 
Binggeli, \& Freeman 2000) and a sample of Virgo and Fornax early-type 
galaxies (Caon et al.\ 1993; D'Onofrio et al.\ 1994).  
The Pearson correlation coefficient between the TGC index and 
luminosity is $-0.82$. 
Given our new understanding of the relationship between the S\'ersic 
index $n$ and the central concentration of luminous mass, the correlation 
between total luminosity and concentration is perhaps not surprising. 
It has, however, to the best of our knowledge, never been shown before
for a sample of Elliptical galaxies.  It can also be derived completely 
independent of any light-profile model (and hence value of $n$), 
although here we have used the TGC index from the best-fitting 
S\'ersic model because of the similarity seen in Figure~\ref{fig6} and 
because we do not have the images for the dwarf galaxy data set. 



To calibrate the luminosity density profiles in Figure~\ref{fig2} requires 
some measure of the central intensity or central surface brightness $\mu_0$ 
which is, 
for an $r^{1/4}$ model, 7.67 mag brighter than the surface  brightness $\mu_e$ 
at one effective radius. Now, the Kormendy (1977) relation for Elliptical 
galaxies tells us that $\mu_e\propto
3\log r_e$, and therefore\footnote{For an $r^{1/4}$  law, $m=\mu_e - 5\log r_e
-3.388$.} the magnitude $M\propto -2\log r_e$.  Thus, assuming $r^{1/4}$
profiles, the more luminous galaxies should possess larger effective radii 
and {\it fainter} central intensities, which equivalently implies lower 
central luminosity densities. 

It must be stressed that we are not referring to the behavior of the 
luminosity density of the core within the central $\sim$arcsecond 
as revealed with HST resolution (Rest et al. 2001, and references
within), but to the properties derived from the global galaxy profile. 
Recent findings show that many Elliptical galaxies possess supermassive 
blackholes at their centers (Kormendy \& Gebhardt 2001 and references within). 
Gravitational slingshots of stars which come to close to the central
massive blackhole, or coalescing massive blackholes from the progenitors 
of a merger may, to varying degrees, evacuate the core of a bulge 
and thereby reduce the original inner light profile (Ebisuzaki, Makino, 
\& Okamura 1991; Makino \& Ebisuzaki 1996; 
Quinlan 1996; Quinlan \& Hernquist 1997; Faber et al.\ 1997; 
Milosavljevi\'c \& Merritt 2001). 
If mergers involve strong gaseous dissipation and
central starbursts this may also modify the nuclear profile 
(Mihos \& Hernquist 1994), as do adiabatic black hole growth models 
(van der Marel 1999a,b, 2001). 
In this paper we are, however, not talking about the very inner density 
of HST-resolved nuclear cusp slopes which may have been 
modified by the central blackhole, but are referring to the global 
galaxy structure as seen with ground based resolution. 

The estimated masses of the central blackholes in bulges have been shown to
positively correlate with the total luminosity of the host galaxy 
(Kormendy 1993; Kormendy \& Richstone 1995), which would imply, assuming 
$r^{1/4}$ profiles, that 
their masses are greater for galaxies with lower central
luminosity densities (again, we are not referring to the modified 
HST resolved cusps). 
Why it should be that galaxies which are globally more 
disperse have greater central blackhole masses must be explained if 
one is to assume that all Elliptical galaxies follow the $r^{1/4}$ law. 
However, this problem is quickly dismissed when one realises that it is
a somewhat artificial problem which was created from the 
simplification of galaxies through the assumption of $r^{1/4}$ profiles.
Not only do a range of light profile shapes exist, as do varying degrees of
concentration, but the central (observed with ground-based telescopes)
surface brightness of all but the brightest Elliptical galaxies actually 
brighten with 
increasing galaxy magnitude (Figure~\ref{fig9})\footnote{There is some 
suggestion of a turn around at $B_T<10$ in Figure~\ref{fig9},  evident amongst the
brightest galaxy members in Figure 6c of Faber et al.\ 1997.} in contradiction 
to the prediction of the Kormendy relation when coupled with the assumption
of $r^{1/4}$ profiles.  That is not to say that the Kormendy relation is 
wrong.  Indeed, the brighter galaxy sample members in Figure~\ref{fig9} 
roughly follow the Kormendy relation, having a slope of $\sim$3 in the 
$\mu_e$--$\log r_e$ plane (Figure~\ref{fig10}). 
(Although, the smaller and fainter galaxies do not follow
the Kormendy relation, but there is certainly no suggestion that $\mu_e$ 
brightens with $r_e$.)  
Figure~\ref{fig9} reveals that the 
mass of the central supermassive blackholes are therefore positively 
correlated with the central luminosity density of the host galaxy as 
derived from the global luminosity profile, with the caveat that the 
very inner light profile cusps have likely been re-shaped by the 
central massive blackhole.  

We wanted to be confident that the trend seen in Figure~\ref{fig9} could not 
be explained by the influence of atmospheric seeing on what might in fact be 
$r^{1/4}$ law profiles which have smaller half-light radii the fainter the 
galaxy magnitudes are.  To explore this we convolved a series of $r^{1/4}$ 
profiles having a range of half-light radii from 5 to 50 arcsec with a 
Gaussian PSF having a FWHM of 2 arcsec, comparable to the worst 
seeing conditions under which the central CCD data was obtained for the
galaxy sample shown in Figure~\ref{fig9}.  We found that the average 
surface brightness within the inner circle of radius 1 arcsec was 
under-estimated by 
$\sim$0.5 mag arcsec$^{-2}$ when $r_e$=5$\arcsec$ and by 
$\sim$0.3 mag arcsec$^{-2}$ when $r_e$=50$\arcsec$.  These differences
where of course even smaller when the values within the inner 2$\arcsec$ radius 
were considered, and the trend seen in Figure~\ref{fig9} did not vary
noticeably when we used the observed data within the inner 2$\arcsec$. 
The effects of seeing are therefore unable to resurrect the possibility 
that the trend seen in Figure~\ref{fig9} is compatible with all
galaxies having $r^{1/4}$ law profiles.

\subsection{Central surface brightness and effective radii}

Khosroshahi, Wadadekar, \& Kembhavi (2000) and Mollenhoff \& Heidt (2000)
have modelled the light profiles from a sample of spiral galaxies 
with a seeing-convolved $r^{1/n}$ bulge and exponential disk model. 
The strongest correlation found between any of the  
structural parameters explored by these authors was between the 
central bulge surface brightness and the bulge shape parameter $n$.  
Khosroshahi et al.\ obtained a linear 
correlation coefficient $r=-0.88$ at a significance level better 
than 99.99\%, and Mollenhoff et al.\ obtained a value of $r=-0.86$. 
Positive correlations between $n$ and $r_e$ were 
also found by these authors. 
Thus, from Section 2, the central luminosity density of spiral galaxy 
bulges (as represented by $\mu_0$) must be positively correlated with 
not only $n$ but also with the bulge concentration of luminous matter. 
One may indeed ask which is the more fundamental connection. 

We show in Figure~\ref{fig11} and Figure~\ref{fig12} the 
correlation between the TGC index and $r_e$ and $\mu_0$ 
for our sample of dwarf Elliptical and Elliptical 
galaxies.  The behavior in Figure~\ref{fig11} is similar to that 
already known between $n$ and 
$r_e$ (e.g.\ Caon et al.\ 1993; Graham et al.\ 1996).  
The correlation coefficient is 0.67, and turns out to be the 
weakest correlation we find between concentration and 
any of the other galaxy parameters presented here. 

In Trujillo et al.\ (2001b) we showed that the range of different
galaxy structures which exists amongst the Elliptical  population cannot be due
to parameter coupling in the $r^{1/n}$  models, and systematically varies with
model-independent quantities  such as effective half-light radius and
luminosity.   We caution, however, that for large values of $n$ the effects of
seeing on the light profile can be substantial at small radii (Trujillo et
al.\ 2001a).  We also note that while the $r^{1/n}$  profiles are good at
describing the global profile shape, they can require modification at some
small inner radii (Jaffe et al.\ 1994; Ferrarese et al.\ 1994; Lauer et al.\
1995;  Faber et al.\ 1997).  The notably bright ($\sim < 15$ mag arcsec$^{-2}$) 
central ($r$=0) surface brightnesses 
expected from the very high values of $n$ are perhaps also unlikely to be
realised.  Either the S\'ersic model is no longer appropriate to describe the
very inner profile, and/or the extremely high inner densities modify the actual
profile.  The presence of (and likely past infall of material into) a central
supermassive BH is also likely to disturb the central cusp slope, as can
mergers and other mechanisms (Makino \& Ebisuzaki 1996; Faber et al.\ 1997; 
Quinlan \& Hernquist 1997; Merritt \& Quinlan 1998).  We therefore note that
Figure~\ref{fig12} (see also Jerjen \& Binggeli 1997), showing the relation 
between the TGC index and the central galaxy surface brightness obtained 
from the best-fitting $r^{1/n}$ model, should be regarded as
somewhat preliminary and a more detailed investigation shall be forthcoming. 
Despite these words of caution, excluding those galaxies with values of $n$
greater than 4, that is, removing those galaxies which may have overly bright
central surface brightness estimates, still resulted in a correlation 
coefficient of -0.88 between the central concentration and central surface
brightness.  
(Using all of the galaxies gave a correlation coefficient $r=-0.94$.) 
We stress again that if all Elliptical galaxies followed the $r^{1/4}$ 
law, then the concentration index would be the same for all galaxies, 
and no correlation would exist between concentration and any of the 
other galaxy parameters; this is similarly the case if all dwarf 
Elliptical galaxies and Spiral galaxy bulges were to possess the same 
universal profile.

\subsection{Velocity dispersion and mass \label{vel}}

Combining the luminosity -- central concentration relation (Figure~\ref{fig8})
with the Faber-Jackson (1976) relation between luminosity and central velocity
dispersion immediately implies that the central velocity dispersion and
therefore mass -- as far as the central velocity dispersion is a measure of
the galaxy mass -- must positively correlate with the galaxy concentration. 
Using the early-type galaxies from Caon et al.\ (1993) and D'Onofrio et 
al.\ (1994) which have available central velocity dispersion 
measurements from Hypercat\footnote{Hypercat can be reached at 
\url{http://www-obs.univ-lyon1.fr/hypercat/.}}, 
Figure~\ref{fig13} shows that the TGC index is strongly 
($r=0.80$ for the Ellipticals) 
correlated with the velocity dispersion, even with the heterogeneous nature 
of the dynamical data.  
Excluding galaxies with central velocity dispersions less than 100 km s$^{-1}$
did not change things appreciably, nor did using the 
central velocity dispersion catalog of McElroy (1995) where we found a value 
of $r=0.82$. 
This is a fundamental result; together with the previous correlations it tells
us that the more massive a galaxy is, the more centrally concentrated it must be.
The virial theorem (and its observational counterpart the Fundamental
Plane: Djorgovski \& Davis 1987; Dressler et al.\ 1987) relates the
luminosity, size, and velocity dispersion (kinetic energy) terms, but does
not explain why greater mass should imply higher central concentration. 
Theories of gravitational collapse and galaxy formation must be able to
explain this.

We speculate here that galaxy concentration may 
provide the physical mechanism for the observed connection 
between the stellar velocity dispersion (mass) of a bulge and the mass of 
its central supermassive BH (Merritt 2000; Ferrarese \& Merritt 2000; 
Gebhardt et al.\ 2000; Merritt \& Ferrarese 2001; Marconi et al.\ 2001; 
Sarzi et al.\ 2001).  
The exact process which explains the reason for the existence of this 
correlation is not yet known, although many theories have been proposed 
(Efstathiou \& Rees 1988; Haehnelt \& Rees 
1993; Ciotti \& Ostriker 1997; 2001; Haiman \& Loeb 1998; Silk \& Rees 1998; 
Blandford 1999; Haehnelt \& Kauffman 2000; Kauffman \& Haehnelt 2000; 
Ostriker 2000; Adams, Graff, \& Richstone 2001).  
Bigger galaxies with higher central concentration and 
stronger potential wells would naturally supply more fuel to their inner
regions.  This could take the form of more efficiently funneling gas to 
build the central accretion disks that likely feed the quasars we observe at
high-redshift (Soltan 1982), perhaps leaving remnants such as the nuclear 
disks we observe today (Rest et al.\ 2001 and references within) which 
would now encircle the inactive heart of what was once a quasar.  

It is not clear why simply having a higher velocity dispersion alone 
can be the fundamental physical mechanism for greater BH masses. 
On the one hand it does imply that more material (mass) is in the 
galaxy to build/feed a blackhole, but we saw in Section~\ref{ozzy} 
that if all galaxies 
were described by the $r^{1/4}$ law then galaxies with greater mass 
(higher velocity dispersion) would have {\it lower} central luminosity 
densities and be more disperse.  
This would require a low-density environment to either favor
the formation of more massive blackholes or to be the product of their 
evolution.  (Once again, we stress that we are not referring to the very
inner nucleus as revealed with HST imaging.)  We propose here that the 
presence of mass itself 
may not be the end of the story.  Figure~\ref{fig13} shows that more 
massive galaxies are more centrally 
concentrated than less massive galaxies.  We suggest that exactly how 
the galaxy mass is distributed/concentrated may be an important factor. 
This could be tested through a photometric campaign which measures the 
concentration in those galaxies with blackhole mass estimates, and
determines the strength and scatter of the correlation between these
two quantities.  
Although, Ferrarese \& Merritt (2001) and Gebhardt et al.\ (2000) 
found that the intrinsic scatter in the BH mass -- velocity 
dispersion diagram is small or negligible (i.e.\ consistent with the 
measurement errors alone) for those galaxies with most reliable blackhole 
mass measurements.   This would then leave no room for 
improvement with a BH mass - galaxy concentration diagram, and would 
suggest that the former is indeed the more fundamental relation.  This 
result may hold firm, although it will be of interest to see 
what happens when more BH mass estimates are obtained and 
refined on several fronts.  Improvements will come with the 
use of non-axisymmetric dynamical models, improved knowledge of 
line broadening mechanisms (Barth et al.\ 2001), and addressing 
concerns that the dynamical effects of some blackholes may occur at 
resolutions lower than presently probed (Qian et al.\ 1995; de Zeeuw 2000).  
Additional corrections for the finite slit width of STIS on the HST 
(Maciejewski \& Binney 2001) may also prove crucial, and significantly 
lower the current blackhole mass estimates.

\section{Conclusions}

We have explored the stability of several different galaxy light 
concentration indices as a function of galaxy exposure depth, or rather,
the number of effective radii sampled.  This analysis has been confined
to one-component stellar systems such as normal Elliptical galaxies, 
dwarf Elliptical galaxies, Spiral galaxy bulges, and exponential disks. 
Our investigation reveals that the ``mean concentration index'', often used
for parametrizing faint and high-redshift galaxies, is a horrendously 
poor estimator of galaxy light concentration and it's use for such a 
task on its own should be abandoned.  To illustrate this claim, we 
have shown that 
this index is unable to distinguish between a giant $r^{1/4}$ Elliptical 
galaxy measured to one effective radii and a pure exponential disk 
measured to three effective radii.  The de Vaucouleurs $C_{31}$ index 
performs notably better but is still heavily dependent on the outer
galaxy radius one reaches.  The central concentration index introduced 
in Trujillo et al.\ (2001b) is shown to be the more stable of the 
indices, changing in value (with increasing galaxy radius) by less than 
10\% once a few effective radii have been sampled, and is more robust 
against measurement errors. 

Given that Elliptical galaxies and bulges are not homologous systems, 
they therefore possess a range of different (light) concentrations. 
The global profile shape, which can be parameterized by 
the value $n$ from the S\'ersic $r^{1/n}$ law, is intimately connected
with the degree of galaxy light (and mass) concentration. 
Which of these two quantities is the more fundamental is not
clear; the latter quantity can however be measured directly from the image
or light-profile, derived independently of any galaxy model. 
For a sample
of dwarf Elliptical galaxies and normal Elliptical galaxies we have 
presented strong correlations between the central concentration index, 
as defined in this 
paper, and the global galaxy parameters: luminosity, effective 
radius, and central surface brightness.   We also present the first 
ever correlation between galaxy concentration and velocity dispersion 
for a sample of Elliptical galaxies, 
showing that the more massive galaxies are more centrally concentrated; 
this should provide a valuable clue into the physics of gravitational 
collapse and galaxy formation.  
Lastly, we speculate and provide a means to test that the central 
concentration of at least luminous matter in Elliptical galaxies, 
that is, how this matter is distributed, may be an 
important quantity regarding the formation of supermassive BHs.

\begin{figure}
\plotone{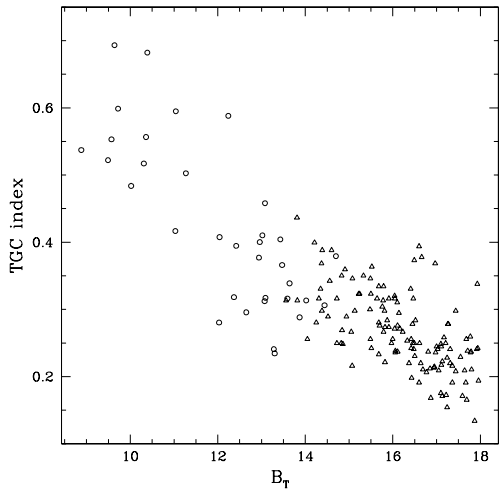}
\caption{The TGC index is plotted against galaxy magnitude 
for a sample of Virgo dwarf Elliptical galaxies (triangles) 
taken from Jerjen, Binggeli, \& Freeman (2000) and a sample 
of early-type Virgo and Fornax galaxies (circles) 
taken from Caon et al.\ (1993) and D'Onofrio et al.\ (1994). 
The TGC index has been derived using equation~\ref{eqTGC2}, and
the magnitudes are the observed (i.e.\ not model-dependent) values. 
All galaxies which could be modelled with a S\'ersic profile along 
the major-axis have been included. 
}
\label{fig8}
\end{figure}
\begin{figure}
\plotone{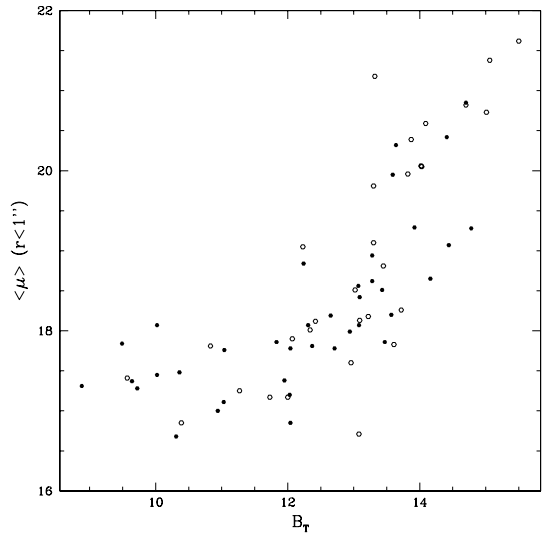}
\caption{The observed (i.e.\ model-independent), central mean surface 
brightness within a circle 
of radius 1 arcsec is plotted against the observed (model-independent)
galaxy magnitude.  Filled circles represent Elliptical galaxies 
and open circles represent S0 galaxies from the complete Virgo and Fornax 
galaxy sample of Caon et al.\ (1993) and D'Onofrio et al.\ (1994) 
observed from the ground.  
}
\label{fig9}
\end{figure}
\begin{figure}
\plotone{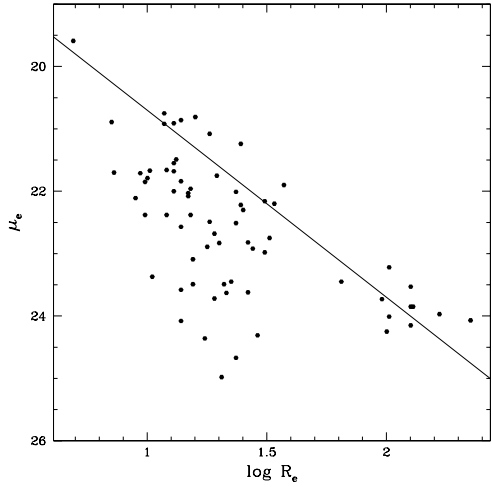}
\caption{The (model-independent) surface brightness $\mu_e$ at 
the (model-independent) effective half-light radius $r_e$ is 
plotted against $r_e$.  The data are from the complete Virgo and Fornax 
galaxy sample studied in Caon et al.\ (1993) and D'Onofrio et al.\ (1994).}
\label{fig10}
\end{figure}
\begin{figure}
\plotone{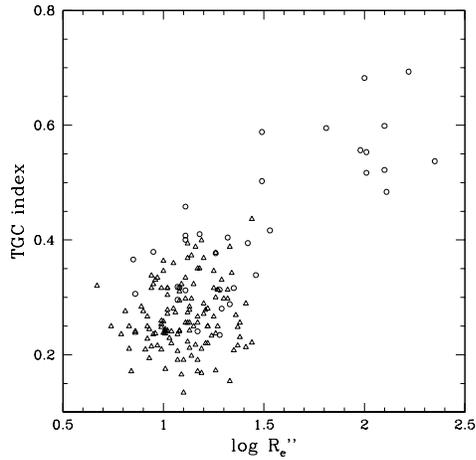}
\caption{
The TGC index shown in Figure~\ref{fig8} is plotted against
each galaxy's (model-independent) effective radii. 
Circles represent the early-type galaxies, and triangles
represent the dwarf Elliptical galaxy sample.
}
\label{fig11}
\end{figure}
\begin{figure}
\plotone{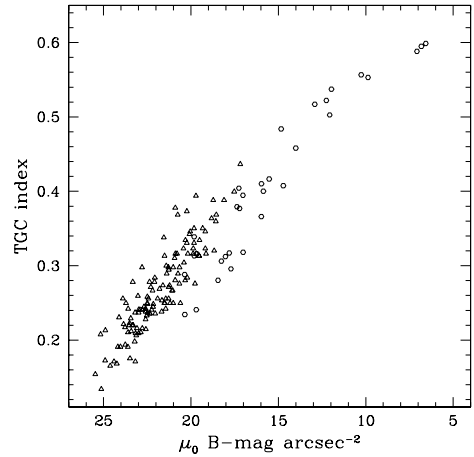}
\caption{
The TGC index shown in Figure~\ref{fig8} has been plotted
against the central surface brightness of each galaxy.
The only two galaxies with values for $n$ greater
than 10 have been excluded (see text for additional comments).
Circles represent the early-type galaxies, and triangles
represent the dwarf Elliptical galaxy sample.
}
\label{fig12}
\end{figure}
\begin{figure}
\plotone{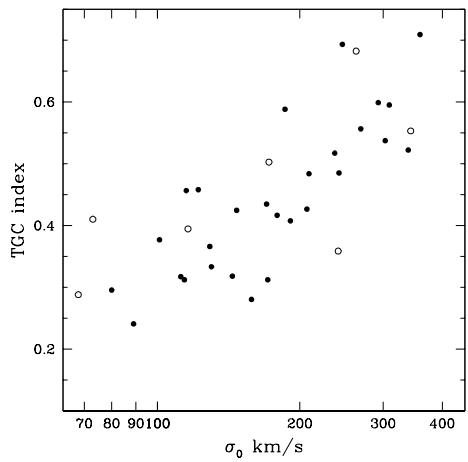}
\caption{
The TGC index shown in Figure~\ref{fig8} has been plotted
against the central velocity dispersion for the early-type 
galaxies with published kinematical data. 
Filled circles represent the Elliptical galaxies, and open circles 
represent S0 galaxies.}
\label{fig13}
\end{figure}

\acknowledgements
We are happy to thank Helmut Jerjen for providing us with the 
dwarf Elliptical data used in Figure~\ref{fig8}-\ref{fig12}. 
We also wish to acknowledge and thank the anonymous referee for 
their suggestions and comments which helped to improve this paper.

\clearpage

\end{document}